\documentclass{article}

\usepackage{enumitem}
\usepackage{multirow}
\usepackage[most]{tcolorbox}
\usepackage{microtype}
\usepackage{tabularx}
\usepackage{graphicx}
\usepackage{subcaption}
\usepackage{booktabs} 

\usepackage{hyperref}

\usepackage[preprint]{icml2026}

\usepackage{amsmath}
\usepackage{amssymb}
\usepackage{mathtools}
\usepackage{amsthm}

\usepackage[capitalize,noabbrev]{cleveref}

\theoremstyle{plain}

\theoremstyle{definition}

\theoremstyle{remark}

\usepackage[textsize=tiny]{todonotes}

\icmltitlerunning{HoliAntiSpoof: Audio LLM for Holistic Speech Anti-Spoofing}

\begin{document}

\twocolumn[
  \icmltitle{HoliAntiSpoof: Audio LLM for Holistic Speech Anti-Spoofing}



  \icmlsetsymbol{equal}{*}

  \begin{icmlauthorlist}
    \icmlauthor{Xuenan Xu}{pjlab}
    \icmlauthor{Yiming Ren}{pjlab}
    \icmlauthor{Liwei Liu}{pjlab}
    \icmlauthor{Wen Wu}{pjlab}
    \icmlauthor{Baoxiang Li}{pjlab}
    \icmlauthor{Chaochao Lu}{pjlab}
    \icmlauthor{Shuai Wang}{nju}
    \icmlauthor{Chao Zhang}{pjlab}
  \end{icmlauthorlist}

  \icmlaffiliation{pjlab}{Shanghai Artificial Intelligence Laboratory}
  \icmlaffiliation{nju}{Nanjing University}

  \icmlcorrespondingauthor{Chao Zhang}{cz277@tsinghua.edu.cn}

  \icmlkeywords{Speech Anti-Spoofing, Speech Deepfake Detection, Multi-modal Large Language Models}

  \vskip 0.3in
]



\printAffiliationsAndNotice{}  

\begin{abstract}
Recent advances in speech synthesis and editing have made speech spoofing increasingly challenging. However, most existing methods treat spoofing as binary classification, overlooking that diverse spoofing techniques manipulate multiple, coupled speech attributes and their semantic effects. 
In this paper, we introduce \textbf{HoliAntiSpoof}, the first audio large language model (ALLM) framework for holistic speech anti-spoofing analysis. HoliAntiSpoof reformulates spoofing analysis as a unified text generation task, enabling joint reasoning over spoofing methods, affected speech attributes, and their semantic impacts. To support semantic-level analysis, we introduce \textbf{DailyTalkEdit}, a new anti-spoofing benchmark that simulates realistic conversational manipulations and provides annotations of semantic influence.
Extensive experiments demonstrate that HoliAntiSpoof outperforms conventional baselines across multiple settings, while preliminary results show that in-context learning further improves out-of-domain generalization.
These findings indicate that ALLMs not only enhance speech spoofing detection performance but also enable interpretable analysis of spoofing behaviors and their semantic effects, pointing towards more trustworthy and explainable speech security.
Data and code are publicly available\footnote{\url{https://github.com/wsntxxn/HoliAntiSpoof}}.

\end{abstract}

\section{Introduction}


Recent advancements in generative models have significantly propelled text-to-speech (TTS) synthesis and speech editing.
Latest models demonstrate the ability to synthesize speech utterances with human-like fidelity and intelligibility~\cite{tan2021survey,du2025cosyvoice3,zhou2025indextts2,peng2024voicecraft}.
While these capabilities offer new applications, they also exacerbate ethical risks, enabling harmful content generation or manipulation of sensitive information.
Consequently, speech anti-spoofing, or speech deepfake detection, has become increasingly critical. 

\begin{figure}
    \centering
    \includegraphics[width=\linewidth]{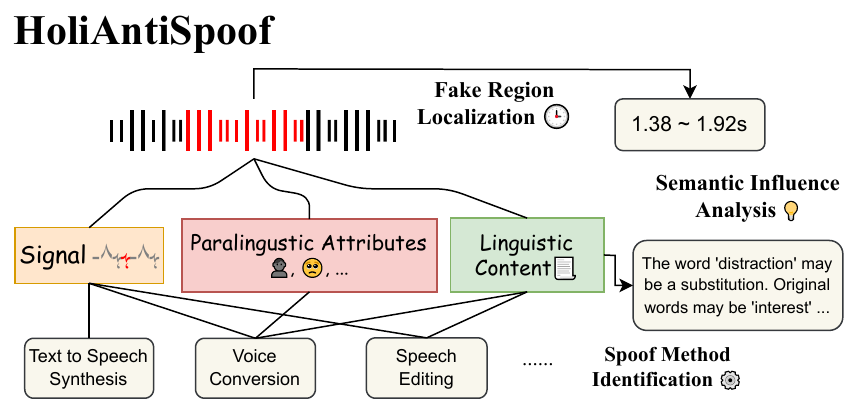}
    \caption{Holistic spoofing analysis incorporates comprehensive speech perception and understanding, to perform multiple subtasks beyond real/fake classification.}
    \label{fig:spoofing_analysis_illustration}
\end{figure}

Traditional anti-spoofing research has predominantly focused on real/fake binary classification.
A back-end classifier typically operates on front-end features~\cite{wang2022investigating,zhang2024audio} or directly on raw waveforms~\cite{tak2021graph,tak2021end,jung2022aasist} to determine whether an utterance is spoofed or to locate spoofed regions within it.
Recent preliminary attempts have explored the adaptation of audio large language models (ALLMs) to spoofing detection~\cite{gu2025allm4add,xie2026interpretable}.  

A binary classification formulation, however, provides an incomplete view of spoofing analysis.
Effective detection requires a comprehensive understanding of speech across multiple levels: 
(i) low-level signal attributes, such as channel information or artifacts introduced during synthesis, 
(ii) paralinguistic attributes, such as speaker identity and emotion, and
(iii) linguistic content conveyed by the text.
As \Cref{fig:spoofing_analysis_illustration} shows, different spoofing methods affect these layers in distinct ways: TTS introduces waveform artifacts at the signal level; voice conversion (VC) can alter paralinguistic attributes, potentially producing unnatural expressions; and neural speech editing modifies portions of real utterances, integrating both signal-level artifacts and linguistic content. 
Consequently, jointly detecting the spoofing method and spoofed regions along with the binary classification enables models to learn anti-spoofing capabilities from the full spectrum of coupled speech information.
In contrast, conventional models may be overfit to specific patterns due to insufficient training data and learning objectives.

Despite its importance, holistic spoofing analysis has been largely unexplored in anti-spoofing research.
Prior anti-spoofing studies have primarily focused on the signal-level authenticity of speech, with limited attention paid to the semantic influence of spoofed utterances.
In practice, manipulations that modify or fabricate the meaning of an utterance can produce far more significant consequences than purely signal-level distortions.
Even subtle edits in sensitive dialogue (\textit{e.g.} changing a ``yes'' to a ``no'' in a financial agreement) can fundamentally invert the intended meaning and lead to substantial real-world impacts, whereas modifications to less critical words (\textit{e.g.}, changing a ``yes'' to a ``yeah'') are often semantically insignificant and less likely to occur in practical spoofing scenarios.
Therefore, analyzing the semantic influence of manipulated words or sentences within context is a crucial yet underexplored component of speech anti-spoofing research.


To overcome these gaps, we present \textbf{HoliAntiSpoof}, the first audio LLM-based system for holistic speech spoofing analysis.
Anti-spoofing inherently requires strong out-of-domain generalization, as generative and editing models continuously evolve and new spoofing methods frequently emerge. 
To this end, we leverage ALLMs for anti-spoofing, which have demonstrated strong capabilities in handling diverse speech understanding tasks within a unified framework. 
Through large-scale pre-training, ALLMs achieve superior performance across tasks, especially on automatic speech recognition (ASR) and speech emotion recognition (SER) \textit{etc}.
Leveraging these strengths, we repurpose an ALLM as a unified speech spoofing analyzer.
Instead of designing separate modules, HoliAntiSpoof integrates all subtasks into a single text prediction task.
Specifically, rich spoofing labels including \textit{real/fake} labels, \textit{spoofing methods}, \textit{spoofed regions}, and potential \textit{semantic influence} are converted into structured annotations and an ALLM is fine-tuned via a standard next-token prediction objective.

Since no existing spoofing datasets provide annotations of the semantic influence of spoofed utterances, we construct two complementary resources.
1) We extend PartialEdit~\cite{zhang2025partialedit}, where at most two words in a single utterance are modified, by annotating the semantic influence of the edits.
2) We propose \textbf{DailyTalkEdit}, derived from DailyTalk~\cite{lee2023dailytalk}, where an entire utterance within a dialogue is replaced by a modified and re-synthesized version.
This simulates realistic conversational manipulations, such as altering a speaker's intended response in multi-turn dialogues, thereby reflecting practical risks in financial, legal, or social contexts.

We train HoliAntiSpoof on a combination of existing spoofing datasets and two newly proposed semantic-oriented datasets. Extensive experiments show that HoliAntiSpoof enables holistic spoofing analysis while consistently outperforming conventional models in spoofing detection. We further investigate the in-context learning (ICL) capabilities of HoliAntiSpoof, demonstrating that providing only a small number of reference examples, without model retraining, substantially improves performance in cross-lingual and out-of-domain spoofing scenarios, highlighting its potential to adapt to unseen data distributions.


Our contributions are summarized as follows:
\begin{itemize}[noitemsep, topsep=0pt, leftmargin=*]
    \item We unify diverse speech anti-spoofing tasks into an ALLM framework, enabling ALLMs to function as holistic spoofing analyzers for the first time.
    \item To incorporate semantic influence analysis into spoofing research, we propose the DailyTalkEdit dataset, facilitating further research.
    \item HoliAntiSpoof achieves state-of-the-art (SOTA) performance on both in-domain and out-of-domain evaluation. Preliminary ICL experiments further show its potential for generalization to new domains.
\end{itemize}


\section{Related Work}

\subsection{Conventional Audio Anti-Spoofing Research}
Audio anti-spoofing, also referred to as audio deepfake detection, aims to defend against malicious audio synthesis and manipulation attacks~\cite{yi2023audio}. Early research primarily focused on designing handcrafted features to capture signal-level artifacts~\cite{todisco2017constant}. With the advent of deep learning, end-to-end models operating directly on raw waveforms have become prevalent~\cite{tak2021end,tak2021graph,jung2022aasist}. More recently, self-supervised learning (SSL) representations, such as HuBERT~\cite{hsu2021hubert} and WavLM~\cite{chen2022wavlm}, have demonstrated strong effectiveness across a wide range of speech tasks, including distinguishing between bona fide and spoofed speech~\cite{wang2022investigating}. Beyond binary utterance-level classification, recent studies have expanded the task to encompass multiple audio domains~\cite{zhang2024svdd,comanducci2024fakemusiccaps,xie2024fakesound} and spoofed region detection within partially manipulated utterances~\cite{zhang2022partialspoof,huang2024detecting}.

Despite these successes, prior works in audio anti-spoofing predominantly focus on the signal-level authenticity of the whole speech utterance or fixed-length segments.
They cannot assess the \textit{semantic influence} of the attack, \textit{i.e.}, how the spoofing operation will potentially alter the speaker's intended meaning.
To bridge this gap, HoliAntiSpoof unifies semantic influence analysis with conventional spoofing detection into the text generation objective of an ALLM.

\begin{figure*}
    \centering
    \includegraphics[width=0.98\linewidth]{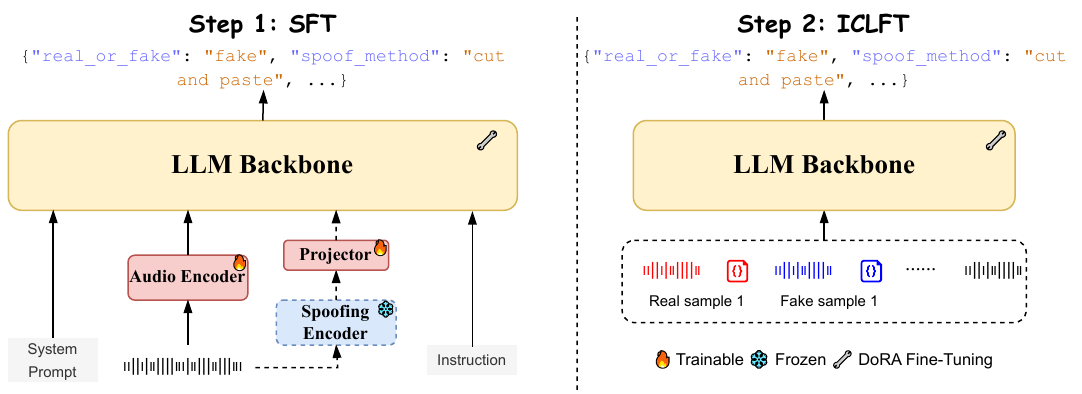}
    \caption{Overview of HoliAntiSpoof. The ALLM is first fine-tuned to generate structured text, including both signal-level and semantic-level spoofing analysis. The incorporation of spoofing-oriented features is explored. Then the LLM is further fine-tuned to enable ICL.}
    \label{fig:main_framework}
\end{figure*}

\subsection{Audio Large Language Models}

Audio LLMs typically align a pre-trained audio encoder with an LLM via a lightweight adaptor, enabling the model to perceive and reason about audio inputs.
Pioneering works like SALMONN~\cite{tang2024salmonn} and Qwen-Audio~\cite{chu2023qwenaudio} have unified diverse speech tasks, such as ASR and SER, into a single instruction-following ALLM.
Recent works explore the integration of discrete audio representations~\cite{ding2025kimi}, unifying understanding and generation~\cite{wu2025step}, and improving the reasoning abilities of ALLMs~\cite{tian2025ualm}.
By scaling up the training data to the magnitude of billions of hours, current ALLMs exhibit remarkable performance in diverse audio understanding tasks.

Recently, researchers have begun to explore the application of ALLMs to audio deepfake detection~\cite{gu2025allm4add,xie2026interpretable}.
\citet{gu2025allm4add} fine-tuned an ALLM for spoofing detection and showed superior performance in data-scarce scenarios, while \citet{xie2026interpretable} applied reinforcement learning with rule-based time- and frequency-domain rewards to improve spoofing detection robustness across diverse audio types.
However, they primarily repurpose the ALLM as a powerful classifier to perform binary classification, underutilizing its potential for holistic spoofing analysis.
Current ALLMs are predominantly pre-trained for semantic extraction and understanding (\textit{e.g.} ASR) and therefore possess weaker sensitivity to low-level acoustic cues such as perceptual quality or subtle synthesis patterns~\cite{wang2025qualispeech}.
To this end, HoliAntiSpoof unifies spoofing detection, localization, method identification, and semantic influence analysis, fully leveraging the strengths of ALLMs to provide holistic spoofing analysis, eliminating the demand for detailed reasoning annotations.


\section{HoliAntiSpoof}

As shown in \Cref{fig:main_framework}, we reformulate the holistic spoofing analysis task to a unified text generation objective to adapt an ALLM for holistic spoofing analysis.
HoliAntiSpoof is first trained by a standard supervised fine-tuning (SFT) and then fine-tuned by an ICL objective.

\subsection{Holistic Spoofing Analysis}
\label{subsec:holistic_spoofing_analysis_definition}

To extend the binary signal-level detection task to holistic analysis, we unify multiple tasks into a single structured text generation objective.
The analysis encompasses the following aspects into a JSON-formatted text:
\begin{itemize}[noitemsep, topsep=0pt, leftmargin=*]
    \item \textbf{Authenticity Classification}: The basic task that classifies whether the input audio is \texttt{real} or \texttt{fake}.
    \item \textbf{Temporal Localization}: For partially spoofed samples, the model predicts the specific time intervals of the spoofed regions.
    \item \textbf{Methodology Identification}: The model categorizes the attack into one of six spoofing techniques: (1) \textit{TTS}, (2) \textit{VC}, (3) \textit{cut and paste}, (4) \textit{speech editing}, (5) \textit{vocoder resynthesis}, and (6) \textit{codec resynthesis}. Detailed explanations are in \Cref{sec:spoof_method_explanation}.
    \item \textbf{Semantic Influence Analysis}: For partially spoofed samples, the model generates a textual description explaining how the modification may influence the original intent, tone, or factual information.
\end{itemize}



\subsection{Model Architecture}

HoliAntiSpoof is built upon a unified ALLM architecture, adopting the pre-trained Qwen2.5-Omni~\cite{xu2025qwen2.5omni} as the initialization backbone.
The overall framework consists of an audio encoder and an LLM backbone.
To assess whether the audio encoder is sufficient for extracting spoofing-relevant representations, we additionally incorporate a spoofing encoder for ablation analysis.

\paragraph{Audio Encoder.}
The audio encoder extracts acoustic features from the raw waveform $\mathcal{A}$.
The input waveform is first converted into log mel-spectrograms, and then processed by convolutional embedding layers and stacked Transformer blocks to produce high-level acoustic embeddings $\mathcal{E}_a \in \mathbb{R}^{l\times d}$ at a temporal resolution of 25 Hz.
To align acoustic representations with the textual modality, a lightweight multi-layer perceptron (MLP) projects $\mathcal{E}_a$ into the embedding space of the LLM backbone.

Although the audio encoder is pre-trained on large-scale data, its training objective primarily emphasizes semantic information, such as speech transcription and audio event recognition.
Low-level acoustic cues related to perceptual quality and synthesis artifacts are often treated as nuisance factors for semantic understanding and may be suppressed during pre-training, despite being highly informative for spoofing detection.
Since holistic spoofing analysis requires both low-level artifact perception and high-level semantic reasoning, we further explore an extension that combines the original audio encoder with an encoder specifically pre-trained for spoofing detection~\cite{ge2025post}.
The spoofing encoder extracts a dense embedding $\mathcal{E}_s \in \mathbb{R}^d$ from the input audio.
After projection through an MLP, $\mathcal{E}_s$ is concatenated with $\mathcal{E}_a$ along the time axis to be fed to the LLM backbone.

\textbf{LLM Backbone.}
The LLM takes multimodal embeddings and generates the target response autoregressively.
We take the thinker part of Qwen2.5-Omni as the backbone.
Given audio embeddings, the model produces structured spoofing analysis outputs in a text generation manner.

\subsection{Training Stage}

\paragraph{SFT.}
The LLM is first fine-tuned by standard SFT.
Formally, the model is trained to maximize the likelihood of the target output $y$ based on the audio embedding $\mathcal{E}_a$:
\begin{equation}
\mathcal{L}_{\mathrm{SFT}} = - \sum_{t=1}^{T} \log p(y_t \mid y_{<t}, \mathcal{E}_{a}),
\end{equation}
where $T$ is the target token number and the system prompt and instruction are omitted for simplicity.

The audio encoder is fully fine-tuned, as spoofing detection may rely on low-level acoustic features that are not fully captured during pre-training.
For the backbone, we adopt DoRA~\cite{liu2024dora} to enable efficient adaptation while preserving the pre-trained model's instruction-following and text generation capabilities.
DoRA decomposes the pre-trained weight into magnitude and direction.
During fine-tuning, the direction is updated via low-rank adaptation, while the magnitude is learned independently, enabling efficient parameter updates with minimal disruption to the pre-trained model:
\begin{equation}
    W' = m\cdot\frac{W_0 + BA}{\Vert W_0+BA\Vert},
\end{equation}
where $W_0 \in \mathbb{R}^{d\times k} $ is the frozen pre-trained weight, $m$ is the trainable magnitude, and $B \in \mathbb{R}^{d\times r}, A \in \mathbb{R}^{r\times k}$ are trainable low-rank matrices.

\paragraph{ICLFT.}
After the standard SFT, we further fine-tune the model in an ICL fine-tuning (ICLFT) paradigm to enable its zero-shot adaptation\footnote{Following \citep{brown2020language}, here ``zero-shot'' means no gradient descent happened at test-time.} capability.
Specifically, we augment the input prompt with a small set of reference examples, including both real and fake audio samples paired with their corresponding structured annotations.
Formally, let $\{(\mathcal{E}_{a_1}, y_{a_1}), \dots, (\mathcal{E}_{a_K}, y_{a_K})\}$ denote $K$ in-context examples.
The training objective is defined as:
\begin{equation*}
\resizebox{\linewidth}{!}{$
\mathcal{L}_{\mathrm{ICLFT}} = 
- \sum_{t=1}^{T} \log p\!\left(
y_t \,\middle|\,
y_{<t},
\underbrace{(\mathcal{E}_{a_1}, y_{a_1}), \dots, (\mathcal{E}_{a_K}, y_{a_K})}_{\text{in-context examples}},
\mathcal{E}_a
\right).
$}
\end{equation*}
By training to condition on a small number of reference examples in ICLFT, the model acquires the ability to infer spoofing-related patterns from context, enabling effective zero-shot generalization to unseen domains and languages with no training at test-time.

\section{Semantic Analysis Data Construction}
\label{sec:data_construction}

\begin{figure*}
    \centering
    \includegraphics[width=0.98\linewidth]{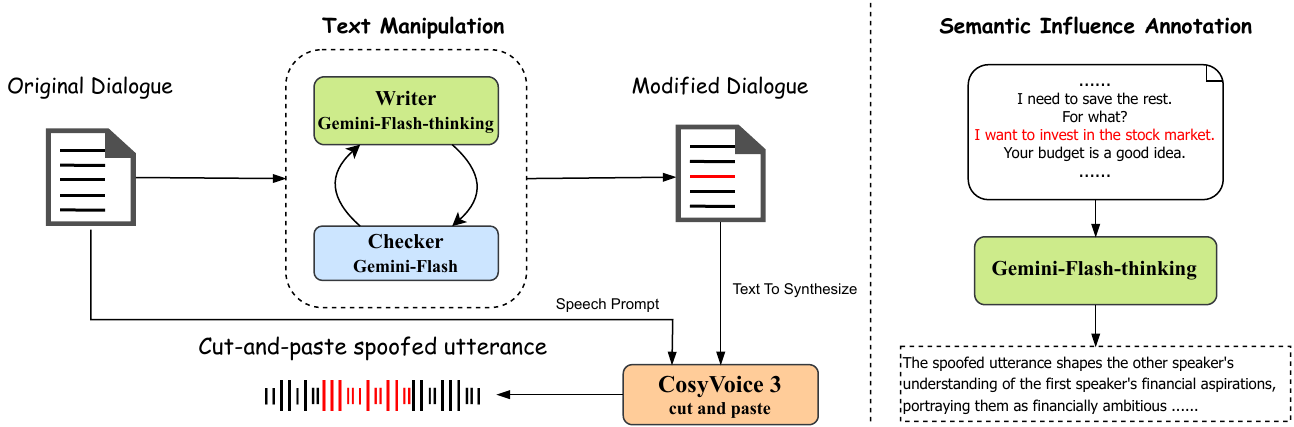}
    \caption{The data construction pipeline of DailyTalkEdit dataset and the annotation of spoofing semantic influence. A dual-agent workflow makes contextual coherent modification of dialogues in DailyTalk for subsequent TTS and cut-and-paste spoofing. Then the spoofed text is fed to Gemini to annotate its potential impact.}
    \label{fig:data_construction}
\end{figure*}

To enable holistic speech deepfake analysis, particularly for understanding the semantic influence of spoofed content, datasets must extend beyond simple binary labels. As existing anti-spoofing datasets lack annotations capturing the semantic effects of spoofing operations, we construct semantic-oriented resources to fill this gap.
We first select \textbf{PartialEdit}~\cite{zhang2025partialedit} as a basis, which modifies one or two words in short single-speaker sentences from the VCTK corpus~\cite{veaux2013voice}.
However, real-world manipulation may also occur in continuous dialogue, where context plays a pivotal role.
To this end, we construct a new dataset named \textbf{DailyTalkEdit}, extending spoofing scenarios to multi-turn dialogues.
Finally, we annotate the semantic influence for both datasets in a unified pipeline.
The data construction workflow is shown in \Cref{fig:data_construction}.

\subsection{DailyTalkEdit Dataset}


Unlike modifying isolated sentences, spoofing within a dialogue requires the manipulated utterance to remain linguistically and semantically coherent with the context, making the attack more realistic. We design an iterative dialogue manipulation workflow incorporating two agents to generate spoofed text.
An advanced zero-shot TTS model is used to synthesize the spoofed utterance to build the dataset.

Directly prompting an LLM to modify a sentence in a dialogue often results in text that contradicts the remaining context.
To generate contextually plausible spoofing samples, we employ a dual-agent workflow involving a \textit{writer} and a \textit{checker}.
\begin{itemize}
    \item \textbf{Writer:} The writer selects a target sentence within the dialogue and generates a corresponding spoofed version.
    It is instructed to alter the semantics (\textit{e.g.}, inverting the intent or changing factual details) to potentially cause negative consequences while preserving contextual coherence.
    We use a reasoning-enhanced Gemini-2.5-Flash for this step, as it requires understanding the whole dialogue and reasoning about semantically plausible modifications.
    \item \textbf{Checker:} Although the writer is instructed to preserve contextual coherence during spoofing, we observe that this requirement is not always satisfied. 
    Thus, we employ another LLM to serve as the \textit{checker}.
    It is responsible for validating contextual coherence and factual consistency between the modified utterance and the original dialogue.
    The non-reasoning Gemini-2.5-Flash is used for this role, as empirical evaluation shows it to be sufficient for coherence verification.
\end{itemize}

The writer iteratively proposes modifications until the checker approves the result or a pre-defined maximum number of iterations $N = 3$ is reached.
Samples that fail to pass the checker within this limit are discarded.
This procedure ensures that the spoofed text is both contextually plausible and challenging to detect.
Starting from 2,541 dialogues in DailyTalk, our procedure produces 2,475 spoofed samples, termed as \textit{DailyTalkEdit} dataset, after filtering.

After generating the modified text, we synthesize the corresponding speech sample for spoofing.
We use a SOTA zero-shot TTS model, CosyVoice-3~\cite{du2025cosyvoice3}, to perform voice cloning.
Specifically, when the original utterance exceeds 2 seconds, it is used as the speech prompt to preserve the speaker's timbre and prosody.
Otherwise, the longest available utterance from the same speaker is selected as the prompt.
The synthesized speech then replaces the original utterance in the dialogue, resulting in a ``cut-and-paste'' style spoofing sample.

\subsection{Semantic Analysis Annotation}

To facilitate training HoliAntiSpoof to analyze the semantic influence of spoofed content, we need to annotate the textual description of how the corresponding spoofing attack may influence the conveyed information. 
We leverage the reasoning capabilities of Gemini-2.5-Flash thinking to generate these descriptions.
Specifically, for each sample in PartialEdit and DailyTalkEdit, we feed the manipulated text or dialogue to the LLM and specify the modified word(s) or sentence in the prompt.
The LLM is required to analyze the impact of the spoofing operation by considering the potential original word or sentence, and generate a concise textual analysis.

\section{Experimental Setup}

\subsection{Datasets}

HoliAntiSpoof is trained on a mixture of existing English anti-spoofing datasets and two newly constructed datasets: the semantic-augmented PartialEdit and DailyTalkEdit.
Detailed descriptions of all datasets are provided in \Cref{subsec:training_data_details}.
Together, these datasets cover the spoofing methods listed in \Cref{subsec:holistic_spoofing_analysis_definition}, spanning both conventional benchmark datasets generated by earlier TTS or VC methods, and recent datasets synthesized using the latest models.
To mitigate data imbalance, we randomly sample at most 50K samples from each dataset during training.

For evaluation, we construct a comprehensive in-domain test set by mixing sampled subsets from each dataset using a real/fake label-stratified strategy, denoted as ``mixed''.
This results in a more balanced evaluation protocol across datasets.
We also report results on the standard benchmark, ASVSpoof2019LA (ASV19.) evaluation set~\cite{nautsch2021asvspoof} for direct comparison with prior works.
In addition, we include four out-of-domain evaluation sets to assess the generalization performance of HoliAntiSpoof on unseen language and speech synthesis models.
Details are provided in \Cref{subsec:evaluation_data_details}.

\subsection{Hyper-Parameters}
During SFT, we optimize the model using AdamW with $\beta=(0.9, 0.95)$ and a weight decay of 0.1.
We use a peak learning rate of $1\times10^{-5}$ with a cosine decay schedule and a warmup ratio of 0.05.
Training is conducted for 20K steps with a batch size of 64.
We apply DoRA to all projection layers in the Transformer backbone, using rank $r=64$, scaling factor $\alpha=128$, and dropout rate 0.05.

During ICLFT, we reduce the learning rate to $5\times10^{-6}$.
The model is trained for 5K steps with a batch size of 32.
The audio encoder is frozen and the DoRA rank $r$ is decreased to 8 to enable a lightweight and stable adaptation.

\subsection{Baselines}
We compare HoliAntiSpoof with the following baseline methods:
\begin{itemize}[noitemsep, topsep=0pt, leftmargin=*]
    \item \textbf{Conventional anti-spoofing models}. We include widely used open-source baselines, including RawNet2~\cite{tak2021end}, RawGAT-ST~\cite{tak2021graph}, AASIST~\cite{jung2022aasist}\footnote{\url{https://github.com/clovaai/aasist}}, and ResNet-Transformer (RT)~\cite{cai2023dku}.
    RawNet2, RawGAT-ST and AASIST operate on raw waveforms, while RT adopts a shallow ResNet-Transformer detector on top of features learned via self-supervised learning (SSL).
    Three SSL features are explored: Wav2Vec2~\cite{baevski2020wav2vec}, HuBERT~\cite{hsu2021hubert}, and WavLM~\cite{chen2022wavlm}.
    These conventional baselines are trained on the same data as HoliAntiSpoof, in a multi-task setting with a shared encoder and multiple task-specific prediction heads.
    Note that RawGAT-ST and AASIST cannot perform spoofing region localization as they produce graph embeddings instead of segment-level embeddings. 
    None of these models can produce semantic influence analysis since they do not include a language decoder.
    \item \textbf{ALLM baselines.} 
    We report results from the latest proprietary ALLM, Gemini-3-Flash, as a reference of the performance of the strongest general ALLM.
    ASVSpoof2019 is excluded from Gemini's evaluation to save inference costs, given the large size of its test set.
\end{itemize}

\begin{table*}[ht]
\centering
\small
\caption{Holistic spoofing analysis performance comparison between HoliAntiSpoof and baseline methods, covering authenticity classification (Auth.), spoofing method identification (Meth.), and spoofing region localization (Loc.). ``F'' and ``T'' indicate whether the SSL backbone is frozen or trainable, respectively. Higher values indicate better performance across all test sets.}
\begin{tabular}{l|cccc|cccc}
\toprule
\multirow{4}{*}{\textbf{Models}} & \multicolumn{4}{c|}{\textbf{In Domain}} & \multicolumn{4}{c}{\textbf{Out Domain}}\\ 
\cmidrule{2-9}
& ASV19. & \multicolumn{3}{c|}{Mixed} & SF-MD. & SpoofCeleb & \multicolumn{2}{c}{HAD} \\
\cmidrule{2-9}
& Auth. & Auth. & Meth. & Loc. & Auth. & Auth. & Auth. & Loc. \\
\midrule
\multicolumn{9}{c}{\textit{Conventional Models}} \\
\midrule
RawNet2 & 77.69 & 91.24 & 90.16 & 53.57 & 88.42 & 68.06 & 5.56 & 37.84 \\
RawGAT-ST & 96.35 & 90.42 & 82.50 & - & 93.69 & 85.70 & 31.86 & -\\
AASIST & 96.29 & 94.29 & \textbf{96.44} & - & 92.11 & 75.91 & 10.49 & - \\
Wav2Vec2 (F) - RT & 79.08 & 70.68 & 65.86 & 41.10 & 69.46 & 41.46 & 47.65 & 37.12 \\
Wav2Vec2 (T) - RT & 79.48 & 91.49 & 90.44 & 53.41 & 91.78 & 41.09 & 22.31 & \underline{42.03} \\
HuBERT (F) - RT & 94.53 & 88.28 & 81.74 & 55.15 & 88.39 & 72.40 & 39.29 & 40.73  \\
HuBERT (T) - RT & 85.32 & 91.14 & 90.99 & \underline{56.37} & 89.30 & 61.63 & 25.80 & 37.36 \\
WavLM (F) - RT & 94.72 & 86.29 & 78.23 & 44.19 & 91.05 & 58.06 & 29.56 & 38.19 \\
WavLM (T) - RT & 74.56 & 88.77 & 90.96 & 54.90 & 91.49 & 63.89 & 14.31 & \textbf{42.96} \\
\midrule
\multicolumn{9}{c}{\textit{ALLM Methods}} \\
\midrule
Gemini-3-Flash & - & 63.39 & 16.10 & 33.26 & 69.49 & 60.73 & 78.59 & 29.55 \\
HoliAntiSpoof & \textbf{96.59} & \textbf{96.16} & \underline{95.12} & \textbf{91.33} & \textbf{97.89} & \textbf{90.36} & \textbf{90.19} & 41.27 \\
\bottomrule
\end{tabular}
\label{tab:main_results}
\end{table*}

\subsection{Metrics}

We evaluate HoliAntiSpoof from multiple perspectives, covering binary spoofing detection and holistic analysis outputs:
\begin{itemize}[noitemsep, topsep=0pt, leftmargin=*]
    \item \textbf{Authenticity classification.}
For real/fake authenticity classification, we report binary classification accuracy (Acc.).
Compared to score-based metrics like equal error rate (EER), Acc. enables comparison with closed-source LLMs that we cannot access probabilities of vocabulary words.
However, we can still calculate EER by extracting token logits at the corresponding position, with details in \Cref{sec:additional_eer_results}.
    \item \textbf{Spoofing method identification.}
For method identification, we use the macro-averaged F$_1$ score.
We merge \textit{TTS synthesis} and \textit{VC} into a single class, as many spoofed samples are generated by hybrid pipelines combining both techniques.
\item
\textbf{Temporal localization.}
For localization of spoofed regions in partially spoofed utterances, we adopt segment-level F$_1$ (Seg-F$_1$)~\cite{mesaros2016metrics} at a temporal resolution of 0.2\,s.
Seg-F$_1$ evaluates localization performance by matching predictions and ground-truth labels on fixed-length segments, making it less sensitive to boundary fragmentation.
\item
\textbf{Semantic influence analysis.}
Evaluating semantic influence analysis is inherently difficult with automatic metrics.
We therefore adopt an LLM-as-a-judge protocol and prompt Gemini-3-Flash to rate the generated analysis on a 1--5 scale along three criteria:
(1) fluency as a standalone text,
(2) correctness in identifying the manipulated word(s) or sentence, and
(3) plausibility of the semantic influence analysis.
To reduce variance, we query the judge three times and report the average score.
\end{itemize}
We apply all four evaluation perspectives on the mixed test set.
For other test sets, we primarily focus on the core task of authenticity classification.
Since HAD~\cite{yi2021had} is dominated by partially spoofed utterances, we additionally report spoofed region localization performance on HAD.

\section{Results}

We first compare HoliAntiSpoof with baseline methods to explore the performance of ALLM on the holistic spoofing analysis task.
Then we investigate key influencing factors of HoliAntiSpoof.

\subsection{Comprehensive Spoofing Analysis Performance}

The comparison between HoliAntiSpoof and conventional baselines is presented in \Cref{tab:main_results}.
On in-domain evaluations, HoliAntiSpoof demonstrates strong and consistent performance across all dimensions.
As a proprietary LLM, Gemini-3-Flash fails to reliably discriminate real and spoofed speech, yielding close-to-chance performance.
Compared with the competitive AASIST, HoliAntiSpoof achieves comparable performance on the widely used ASVSpoof2019.
On the comprehensive mixed in-domain evaluation set, HoliAntiSpoof outperforms all conventional methods and attains spoofing method classification results that are close to the best-performing AASIST.
For ResNet-Transformer baselines, making the SSL feature extractor trainable generally improves authenticity classification performance on the mixed test set but degrades performance on ASVSpoof2019, indicating limited robustness across domains.
In contrast, HoliAntiSpoof maintains high detection accuracy on both evaluation sets.
Regarding temporal localization, HoliAntiSpoof substantially outperforms conventional approaches, achieving a Seg-F$_1$ score above 90, whereas all conventional models remain below 60.

On out-of-domain evaluations, HoliAntiSpoof demonstrates more pronounced advantages over conventional baselines.
While several conventional models generalize well on SF-MD., their performance degrades substantially on other out-of-domain test sets.
In particular, conventional approaches reveal a trade-off between in-domain and out-of-domain performance: for example, although AASIST outperforms RawGAT-ST on the in-domain mixed dataset, it falls behind RawGAT-ST when transferred to unseen out-of-domain test sets.
In contrast, HoliAntiSpoof maintains consistently high spoofing detection accuracy across both in-domain and out-of-domain evaluations, suggesting stronger robustness to domain shifts.
We hypothesize that this improved generalization may be attributed to the large model capacity and transferable representations. 

\begin{table*}[ht]
\centering
\caption{Ablation and analysis results regarding the effect of the data format, ICL with 3 pairs of in-context examples, and the incorporation of spoofing-oriented features. Higher values indicate better performance across all test sets.}
\begin{tabular}{l|cccc|ccc}
\toprule
\multirow{4}{*}{\textbf{Models}} & \multicolumn{4}{c|}{\textbf{In Domain}} & \multicolumn{3}{c}{\textbf{Out Domain}}\\ 
\cmidrule{2-8}
 & \multicolumn{4}{c|}{Mixed} & SpoofCeleb & \multicolumn{2}{c}{HAD} \\
\cmidrule{2-8}
& Auth. & Meth. & Loc. & Sem. & Auth. & Auth. & Loc. \\
\midrule
HoliAntiSpoof & 96.16 & \textbf{95.12} & 91.33 & 3.51 & 90.36 & 90.19 & 41.27 \\
\hspace{1em} w. ICLFT & 96.08 & 94.93 & 90.19 & \textbf{3.64} & 91.24 & \textbf{93.92} & \textbf{79.06} \\
\hspace{1em} w. spoof-feature & \textbf{98.36} & 94.37 & 90.82 & 3.59 & \textbf{94.82} & 88.26 & 46.39 \\
DoRA $r=16$ & 96.34 & 94.53 & 91.04 & 3.55 & 91.90 & 93.92 & 38.68 \\
DoRA $r=256$ & 95.74 & 94.69 & 90.24 & 3.63 & 93.83 & 95.61 & 43.81 \\
\hspace{1em} Auth. Only & 96.68 & - & - & - & 94.96 & 68.20 & - \\ 
\bottomrule
\end{tabular}
\label{tab:analysis_results}
\end{table*}

Analysis of out-of-domain results shows that HoliAntiSpoof exhibits trends similar to those observed in conventional models. In particular, the spoofing method remains the dominant factor influencing cross-domain generalization. For example, performance remains high on SF-MD., as it employs the same generation models as SF-BD., which is included in the training data, despite the target language being unseen during training.
Conversely, when the spoofing method is unseen, performance degrades substantially, as observed on SpoofCeleb.
Moreover, when both the language and spoofing method are unseen, the degradation becomes more pronounced, as reflected by the results on HAD.

\subsection{Ablation Studies and Analysis}

We further analyze the impacts of different components on HoliAntiSpoof by investigating the following questions: 
\textbf{1)} Is holistic spoofing analysis beneficial for spoofing detection? \textbf{2)} Can ICL help improve the generalization performance of HoliAntiSpoof? \textbf{3)} Are spoofing-oriented features complementary to the original audio features for HoliAntiSpoof?

\subsubsection{Influence of Unified Spoofing Target}
\label{subsubsec:influence_data_format}


We examine the effect of the unified spoofing target on basic spoofing detection performance.
Following \citep{gu2025allm4add}, we simplify the learning objective to authenticity classification by training the model to generate a single-token label (\texttt{real} or \texttt{fake}), enabling a direct comparison between HoliAntiSpoof and an LLM used purely as a classifier.
Without the holistic analysis objective, the model achieves comparable performance on in-domain test sets but exhibits a substantial performance drop on the out-of-domain HAD dataset.
Moreover, as shown by the EER results in \Cref{tab:additional_eer_results}, HoliAntiSpoof also outperforms the authenticity-only variant on SpoofCeleb under the optimal decision threshold.
These results indicate that the holistic spoofing analysis objective improves generalization by encouraging the model to learn and exploit inherent correlations across tasks.

\subsubsection{Incorporation of In-Context Learning}


We further examine whether ICLFT improves cross-domain generalization.
For in-domain evaluation, only the target audio is provided as input, whereas for out-of-domain evaluation, we additionally prepend 3 pairs of in-context examples (without gradient descent) in the prompt.
As shown in Table \ref{tab:analysis_results}, ICLFT (w. ICLFT) causes only a marginal performance drop on the in-domain mixed test set, suggesting that the model's core spoofing analysis capability is largely preserved after fine-tuning.
On out-of-domain test sets, ICLFT brings a substantial improvement on HAD, which mainly consists of partially spoofed samples generated by cut-and-paste manipulation.
The improvement suggests that ICLFT enhances the model's ability to leverage in-context examples for domain adaptation.
However, on SpoofCeleb, where spoofing patterns are more diverse, ICLFT provides limited gains, suggesting that unlocking the full potential of ICL for spoofing generalization remains an open challenge.

\subsubsection{Benefits of Incorporating Spoofing-Oriented Features}
Finally, we study the effect of audio representations.
As shown in Table \ref{tab:analysis_results}, adding spoofing-oriented features (w. spoof-feature) substantially improves authenticity classification on most datasets, with spoofing method identification and region localization performance slightly degraded.
We also vary the DoRA rank $r$ to compare the effect of acoustic representations and the LLM adaptation capacity: compared with increasing $r$, incorporating spoofing-oriented features yields markedly larger gains in authenticity classification, while increasing $r$ improves the semantic influence score.
The spoofing detection benefits more from representative features, while semantic analysis and reasoning benefit more from larger LLM adaptation capacity.
This implies that the pre-trained ALLM is effective at capturing high-level semantic content in audio but lacks spoofing-sensitive low-level acoustic features for authenticity classification.
Overall, these results confirm that spoofing-oriented representations are complementary to the original audio features for holistic spoofing analysis.


\section{Conclusion}
In this work, we present HoliAntiSpoof, the first holistic speech spoofing analysis framework based on ALLMs. HoliAntiSpoof extends conventional binary real/fake classification to a unified analysis that encompasses spoofing method identification, spoofed region localization, and semantic influence analysis, all formulated as a structured text generation task. To support semantic reasoning, we augment PartialEdit with semantic annotations and introduce DailyTalkEdit, a dialogue-level dataset that simulates realistic conversational spoofing scenarios. Experimental results demonstrate that HoliAntiSpoof substantially outperforms conventional baselines in both in-domain and out-of-domain evaluations, while incorporating spoofing-oriented features further improves authenticity classification accuracy. Finally, preliminary exploration of ICL suggests its potential for adapting ALLMs to unseen domains with only a few examples, motivating future work to fully exploit its capabilities.



\newpage
\section*{Impact Statement}

The rapid advancement of AIGC has made high-fidelity speech deepfakes a significant threat to personal privacy, financial security, and social stability.
By unifying signal-level detection with semantic influence analysis, this work provides a more comprehensive and interpretable tool for the general public to identify and understand malicious audio manipulations. 
While the primary goal of this research is defensive, we acknowledge that the methodology used to analyze spoofing mechanisms could theoretically be studied by adversaries to develop more sophisticated and indistinguishable attacks.
However, we believe that the benefits of providing a robust, generalized, and interpretable defense system far outweigh these risks.

\nocite{langley00}

\bibliography{example_paper}
\bibliographystyle{icml2026}

\newpage
\appendix
\onecolumn

\section{Spoofing Methodology Explanation}
\label{sec:spoof_method_explanation}

\begin{itemize}
    \item \textbf{Text-to-Speech Synthesis (TTS):} The whole speech utterance is generated by TTS models from the textual input.
    \item \textbf{Voice Conversion (VC):} The whole speech is a modified version of a source speaker's utterance, with the timbre and prosody altered while the original linguistic content preserved.
    \item \textbf{Cut and Paste (CaP):} Segments from different recordings of the same speaker to are concatenated to form new sentences, often used to alter semantic meaning without generative models.
    \item \textbf{Speech Editing (SE):} Generative models (codec-based models~\cite{peng2024voicecraft} or diffusion models~\cite{le2023voicebox}) are used to modify specific regions of an utterance to alter keywords while maintaining contextual coherence.
    \item \textbf{Vocoder Resynthesis (VR):} Acoustic features (e.g., mel-spectrograms) of real speech utterances are extracted and resynthesized to raw waveforms using a neural vocoder, introducing specific generation artifacts.
    \item \textbf{Codec Resynthesis (CR):} Real speech utterances are compressed and decompressed through neural audio codecs, where quantization artifacts may be introduced.
\end{itemize}

\section{Data Details}

\subsection{Training}
\label{subsec:training_data_details}

\begin{table}[ht]
    \centering
    \caption{Details of HoliAntiSpoof training set.}
    \begin{tabular}{ccc}
    \toprule
    Dataset & Spoof Method & \# Samples \\
    \midrule
    ASVSpoof2019 (ASV19.)~\cite{nautsch2021asvspoof} & TTS, VC & 25,380 \\
    PartialSpoof~\cite{zhang2022partialspoof} & CaP & 25,380 \\
    SINE~\cite{huang2024detecting} & CaP, SE & 104,942 \\
    WaveFake~\cite{frank2021wavefake} & VR & 71,883 \\
    CodecFake~\cite{wu2024codecfake} & CR & 17,267 \\
    PartialEdit~\cite{zhang2025partialedit} & CaP, SE & 63,204 \\
    VCTK~\cite{veaux2013voice} & / & 7,730 \\
    LJSpeech~\cite{keith2017ljspeech} & / & 10,269 \\
    EmoFake~\cite{yan2024emofake} & VC & 27,300 \\
    SpeechFake~\cite{huang2025speechfake}-en & TTS, VC & 458,245 \\
    DailyTalkEdit & CaP & 3,600 \\
    FakeOrReal~\cite{reimao2019dataset} & Unknown & 53,868 \\
    InTheWild~\cite{muller2022does} & Unknown & 31,779 \\
    \bottomrule
    \end{tabular}
    \label{tab:training_data_details}
\end{table}

HoliAntiSpoof is trained on a mixture of diverse English anti-spoofing datasets.
Details of each dataset is listed in \Cref{tab:training_data_details}.
For SpeechFake, English samples for the bilingual and multilingual subsets are used for training.
For VCTK and LJSpeech, we utilize disjoint speaker sets for training and testing; consequently, the mixed training set incorporates only a subset of these two corpora.

Among these datasets, the semantic influence of the spoofed content is only available for PartialEdit and DailyTalkEdit.
The spoofing methods for FakeOrReal and InTheWild datasets are not available.
To accommodate these varying levels of annotation granularity, we use task-specific prompts.
This strategy prevents the model from encountering inconsistent target formats during training.

\subsection{Evaluation}
\label{subsec:evaluation_data_details}

\begin{table}[ht]
    \centering
    \caption{Details of HoliAntiSpoof evaluation dataset.}
    \begin{tabular}{ccccc}
    \toprule
     & Dataset & Language & Spoof Method & \# Samples \\
    \midrule
    \multirow{11}{*}{\shortstack{Mixed \\ In Domain}} & ASVSpoof2019 (ASV19.)~\cite{nautsch2021asvspoof} & \multirow{11}{*}{en} & TTS, VC & 2,000 \\
    & PartialSpoof~\cite{zhang2022partialspoof} & & CaP & 2,000 \\
    & SINE~\cite{huang2024detecting} & & CaP, SE & 2,000 \\
    & WaveFake~\cite{frank2021wavefake} & & VR & 2,000 \\
    & CodecFake~\cite{wu2024codecfake} &  & CR & 2,000 \\
    & PartialEdit~\cite{zhang2025partialedit} & & CaP, SE & 2,000 \\
    & VCTK~\cite{veaux2013voice} & & / & 2,000 \\
    & LJSpeech~\cite{keith2017ljspeech} & & / & 1,365 \\
    & EmoFake~\cite{yan2024emofake} & & VC & 2,000 \\
    & SpeechFake~\cite{huang2025speechfake}-en & & TTS, VC, VR & 10,000 \\
    & DailyTalkEdit & & CaP & 950 \\
    \midrule
    \multirow{3}{*}{Out Domain} & SpeechFake MultiLingual (SF-MD.)~\cite{huang2025speechfake} & 11 languages & TTS, VC & 11,000 \\
     & SpoofCeleb~\cite{jung2025spoofceleb} & multilingual & TTS & 10,000 \\
     & HAD~\cite{yi2021had} & zh & TTS, CaP & 9,072 \\
    \bottomrule
    \end{tabular}
    \label{tab:evaluation_data_details}
\end{table}

Details of the data used for evaluation are listed in \Cref{tab:evaluation_data_details}.
We sample from each datasets using a real/fake label-stratified strategy to construct the mixed in-domain test set.
For most datasets, the sampling number is 2,000.
Since SpeechFake-en comprises plenty of latest TTS, VC and vocoder models, we increase its sample size to 10,000 to ensure that each underlying generative model is adequately represented. 
As the available samples for \textit{LJSpeech} and \textit{DailyTalkEdit} test sets are fewer than 2,000, sampling is not needed and all available data are used for evaluation.

\section{Additional EER Results}
\label{sec:additional_eer_results}


\begin{table*}[ht]
\centering
\small
\caption{Additional anti-spoofing metrics using equal under rate (EER). }
\begin{tabular}{l|cc|cc}
\toprule
\multirow{2}{*}{\textbf{Models}} & \multicolumn{2}{c|}{\textbf{In Domain}} & \multicolumn{2}{c}{\textbf{Out Domain}}\\ 
\cmidrule{2-5}
& ASV19. & Mixed & SF-MD. & SpoofCeleb \\
\midrule
\multicolumn{5}{c}{\textit{Conventional Models}} \\
\midrule
RawNet2 & 6.58 & 8.67 & 6.36 & 31.21 \\
RawGAT-ST & 2.86 & 6.65 & 6.08 & 21.20 \\
AASIST & 2.65 & 4.60 & 7.41 & 16.80 \\
Wav2Vec2 (F) - RT & 13.51 & 24.89 & 19.57 & 44.19 \\
Wav2Vec2 (T) - RT & 13.96 & 8.04 & 9.79 & 34.00 \\
HuBERT (F) - RT & 5.56 & 10.61 & 7.71 & 28.00 \\
HuBERT (T) - RT & 12.48 & 8.70 & 12.43 & 27.00 \\
WavLM (F) - RT & 5.97 & 12.70 & 6.95 & 32.59 \\
WavLM (T) - RT & 14.47 & 9.62 & 5.77 & 25.40 \\
\midrule
\multicolumn{5}{c}{\textit{ALLM Methods}} \\
\midrule
HoliAntiSpoof-Auth.-Only & \textbf{1.10} & \textbf{3.06} & \underline{0.68} & \underline{10.03} \\ 
HoliAntiSpoof & \underline{1.11} & \underline{3.48} & \textbf{0.39} & \textbf{8.52} \\
\bottomrule
\end{tabular}
\label{tab:additional_eer_results}
\end{table*}

\Cref{tab:additional_eer_results} presents the binary authenticity classification performance of HoliAntiSpoof and baseline systems measured in EER.
Following \citet{gu2025allm4add}, we compute EER based on the normalized probability of predicting \texttt{real}:
\begin{equation}
    p(\text{real}) = \frac{\exp(s_{\text{real}})}{\exp(s_{\text{real}}) + \exp(s_{\text{fake}})},
\end{equation}
where $s_{\text{real}}$ and $s_{\text{fake}}$ denote the logits of the tokens \texttt{real} and \texttt{fake} at the position immediately after the prefix ``\texttt{\{"real\_or\_fake": }''.
Since logits of each token are not available from Gemini-3-Flash, it is not included for comparison.
The EER results are largely consistent with the accuracy trends in \Cref{tab:main_results}, further validating the superiority of HoliAntiSpoof. 
In in-domain evaluations, HoliAntiSpoof-Auth.-Only and HoliAntiSpoof demonstrate comparable performance, both significantly outperforming conventional models.
On out-of-domain SF-MD. and SpoofCeleb, HoliAntiSpoof-Auth.-Only achieves higher classification accuracy, but lags bebind HoliAntiSpoof in terms of EER.
This suggests that holistic spoofing analysis training enhances HoliAntiSpoof's discriminative ability, leading to superior robustness in unseen domains when evaluated under an optimal threshold.
Consequently, the classification accuracy of HoliAntiSpoof can be optimized by employing a calibration set from the target domain to refine the decision threshold.

\section{Prompts}

\subsection{Semantic Influence Evaluation}
We present below the prompt used for evaluating the quality of semantic influence analysis on spoofed content.
The quality is judged from three perspectives.

\begin{tcolorbox}[
    colback=gray!5!white, 
    colframe=gray!75!black, 
    title=LLM Judge Prompt for Semantic Influence Analysis,
    fonttitle=\bfseries,
    fontupper=\ttfamily\scriptsize
]
\#\# Role

You are a senior expert in audio spoofing detection and content safety assessment.
Your job is to score the output quality of an "audio spoofing detection model".

\medskip

\#\# Task

Compare the [Original Content] and [Spoofed Content] to check if the [Model Analysis Result] accurately identified the tampering behavior and its impact.

\medskip

\#\# Original Content

[The entire speech transcription]

\medskip

\#\# Spoofed Content

[Original word/sentence] -$>$ [New word/sentence]

\medskip

\#\# Model Analysis Result

[Model output]

\medskip

\#\# Scoring Rubric (1-5 Scale)

5 (Perfect):

Precise Localization: Accurately identifies the tampered position.

Deep Analysis: Logical and rigorous; accurately identifies the profound impact of tampering on semantic integrity, character personas, or conversational context.

Fluent Expression: Smooth phrasing with no grammatical errors.

\medskip

4 (Excellent):

Precise Localization: Accurately identifies the tampered position.

General Analysis: Identifies the tampering facts, but the analysis of contextual impact lacks depth or contains minor logical flaws.

Fluent Expression: Smooth phrasing.

\medskip

3 (Passing):

Precise Localization: Accurately identifies the tampered position.

Erroneous Analysis: Correct location identified, but the analysis of tampering intent or semantic changes is entirely wrong or contradicts the facts.

Fluent Expression: Smooth phrasing.

\medskip

2 (Poor):

Incorrect Localization: Fails to identify the correct tampered position (significant deviation) or identifies the wrong segment.

Fluent Expression: Phrasing is smooth despite the incorrect content.

\medskip

1 (Invalid):

Total Failure: Fails to identify tampering, or the output is irrelevant to the task (e.g., gibberish, completely off-topic).

\medskip

\#\# Output Rules

Judge strictly based on the scoring rubric.

Output the score directly; do not include any Markdown tags or other explanatory text.
\end{tcolorbox}

\subsection{DailyTalkEdit Data Curation}
As \Cref{fig:data_construction} shows, the dialogue in the DailyTalk dataset is modified by a dual-agent workflow.
The writer selects a target sentence to modify while the checker validates whether the modified sentence is contextually coherent without obvious contradictions.
Corresponding prompts are demonstrated below.
The workflow is designed to alter critical information in the dialogue as much as possible, while keeping the modified dialogue logically consistent and reasonable.

\begin{tcolorbox}[
    colback=gray!5!white, 
    colframe=gray!75!black, 
    title=Prompt for Writer,
    fonttitle=\bfseries,
    fontupper=\ttfamily\scriptsize
]
You are simulating a malicious spoof attack on a dialogue. Your task is to:

1. Read the entire dialogue carefully, paying special attention to the context before and after each utterance.

2. Identify the utterance that, if maliciously modified, could cause the most harm, confusion, or negative consequences.

3. Modify the utterance in a way that is plausible but dangerous - the kind of modification that a spoof attack would make.
 
\medskip

CAUTION:

1. The modified sentence MUST blend in naturally with the rest of the dialogue and not stand out due to unusual length. This is essential for a realistic spoof attack. 

2. The modified utterance MUST maintain logical coherence with the surrounding dialogue. Even though you are changing critical information, the subsequent utterances in the dialogue should still make sense in context. Avoid creating obvious contradictions that would immediately reveal the manipulation.

\medskip

The dialogue is given below. Each line is prefixed with an index in square brackets, which indicates the utterance index in the dialogue:

[Dialogue]

Now, follow these rules strictly:

- You MUST select exactly ONE utterance to modify that has the potential for maximum harm or negative impact.

- You MUST output a JSON object with the following format, and NOTHING else:

\{
  
\hspace*{1em}``target\_utterance\_idx'': $<$int index of the selected utterance$>$,

\hspace*{1em}``modified\_text'': $<$the maliciously modified utterance text$>$

\}

Do not add any explanation or extra text outside this JSON.
\end{tcolorbox}

\begin{tcolorbox}[
    colback=gray!5!white, 
    colframe=gray!75!black, 
    title=Prompt for Checker,
    fonttitle=\bfseries,
    fontupper=\ttfamily\scriptsize
]
You are a dialogue coherence checker. Your task is to evaluate whether a modified utterance in a dialogue maintains logical coherence with the surrounding context.
\medskip
A modification PASSES if:

1. The modified utterance fits naturally within the dialogue flow

2. Subsequent utterances in the dialogue still make logical sense after the modification

3. There are no obvious contradictions that would immediately reveal the modification as suspicious
\medskip

A modification FAILS if:

1. Subsequent utterances contradict or become nonsensical after the modification

2. There are clear logical contradictions between the modified utterance and surrounding context
\medskip

Note: The modification is INTENDED to be malicious (changing critical information). Your job is to check if it's done in a way that maintains surface-level coherence - a successful deepfake attack should be subtle enough that the dialogue still flows naturally.
\medskip

ORIGINAL DIALOGUE:

[Original Dialogue]

MODIFIED DIALOGUE (utterance [Modified Utterance Index] has been modified):

[Modified Dialogue]

\medskip

Evaluate whether the modification maintains logical coherence. Pay special attention to:

- Whether utterances after [Modified Utterance Index] still make sense in context

- Whether there are obvious contradictions that would immediately reveal the modification

- Whether the dialogue flow remains natural and coherent

\medskip

You MUST output a JSON object with the following format, and NOTHING else:

\{

\hspace{1em}  ``passed'': $<$true or false$>$,

\hspace{1em}  ``reason'': $<$brief explanation of why it passed or failed$>$

\}

Do not add any explanation or extra text outside this JSON.
\end{tcolorbox}



\end{document}